# KLMAT: A Kernel Least Mean Absolute Third algorithm

Lu Lu, Haiquan Zhao, Badong Chen

**Abstract.** In this paper, a kernel least mean absolute third (KLMAT) algorithm is developed for adaptive prediction. Combining the benefits of the kernel method and the least mean absolute third (LMAT) algorithm, the proposed KLMAT algorithm performs robustly against noise with different probability densities. To further enhance the convergence rate of the KLMAT algorithm, a variable step-size version (VSS-KLMAT algorithm) is proposed based on a Lorentzian function. Moreover, the stability and convergence property of the proposed algorithms are analyzed. Simulation results in the context of time series prediction demonstrate that the effectiveness of proposed algorithms.

## 1  Introduction

The least mean square (LMS) algorithm is most popular adaptive filter because of its low-cost and robustness. However, the convergence rate of LMS algorithm is slow when the measurement noise is added to the unknown system. For this reason, the LMAT algorithm [1] and its family are introduced based on high-order-error-power (HOEP) criterion [2-3]. Particularly, the LMAT algorithm achieves the robust performance against the noises with different probability densities. Moreover, its cost function is convex and therefore has no local minima.

Recently, kernel methods have become rather popular and have been successfully applied to diverse fields [4]. Due to their universal modeling capability and convex cost, kernel adaptive filters (KAFs) are attracting renewed attention. The main idea of the KAF is that recast the input data into a high-dimensional feature space via a reproducing kernel Hilbert space (RKHS). Then, the linear adaptive filter is applied in the feature space. Based on these considerations, several kernel adaptive algorithms were proposed [5-6]. Among these KAFs, the KLMS algorithm has been widely used as a nonlinear system modeling tool with considerable success [7]. Moreover, many other kernel adaptive algorithms ad-

L. Lu and H. Zhao*
Key Laboratory of Magnetic Suspension Technology and Maglev Vehicle, Ministry of Education, and School of Electrical Engineering, Southwest Jiaotong University, Chengdu, China
e-mail: lulu@my.swjtu.edu.cn, hqzhao_swjtu@126.com

B. Chen
School of Electronic and Information Engineering, Xi'an Jiaotong University, Xi'an, China.
e-mail: chenbd@mail.xjtu.edu.cn

dressing specific problems were also proposed, such as those proposed in [8-9] and references therein.

To improve the robustness of kernel adaptive algorithm against interferences with various probability densities, in this letter, we proposed a kernel LMAT (KLMAT) algorithm. Then, a variable step-size (VSS) adjustment process based on *Lorentzian function* is incorporated into the KLMAT algorithm, resulting in a new VSS-KLMAT algorithm. The stability and convergence property analysis of the proposed algorithms are performed. Simulation results show that the proposed algorithms in general outperform the existing algorithms for time series prediction under the different measurement noises.

This paper is structured as follows: in Section 2, a KLMAT algorithm and a VSS-KLMAT algorithm are proposed based on kernel method. In Section 3, convergence property analysis of the proposed algorithm is performed. The comparative results of the proposed algorithms are shown in Section 4 and conclusions are drawn in Section 5.

## 2 Proposed algorithms

### 2.1 KLMAT algorithm

The kernel methods are a powerful nonparametric modelling tool. The key to kernel method is transforming input data (input space $\mathbb{U}$) into a high-dimensional feature space $\mathbb{F}$ by using a certain nonlinear mapping. As a result, the inner product operations in the linear adaptive filters are translated to the calculation of a kernel function $\kappa$ in the feature space without knowing the exact nonlinear mapping. To enhance the performance of KLMS algorithm, the LMAT algorithm is first applied in RKHS to generate a KLMAT algorithm. The input data of LMAT algorithm $\mathbf{u}(n)$ are transformed into RKHS as $\boldsymbol{\varphi}(\mathbf{u}(n))$. For convenience, $\boldsymbol{\varphi}(\mathbf{u}(n))$ is replaced by $\boldsymbol{\varphi}(n)$ throughout this paper. The weight vector $\mathbf{w}(n)$ in feature space is defined as $\boldsymbol{\Omega}(n)$. Define $\boldsymbol{\Omega}(1)=\mathbf{0}$ and the error signal $e(n)=d(n)-\boldsymbol{\Omega}^T(n)\boldsymbol{\varphi}(n)$, where $d(n)$ denotes the desired signal. The cost function of KLMAT algorithm is defined as

$$J(n) = |e(n)|^3. \tag{1}$$

Minimizing the instantaneous third power of absolute error value, the adaptation of KLMAT algorithm in RKHS can be expressed as

$$\boldsymbol{\Omega}(n+1) = \boldsymbol{\Omega}(n) - \frac{\mu}{3}\nabla_{\boldsymbol{\Omega}(n)}J(n) \tag{2}$$

where $\nabla_{\boldsymbol{\Omega}(n)}J(n) = -3e^2(n)sign\{e(n)\}\boldsymbol{\varphi}(n)$ is the gradient vector and $\mu$ is the step size (learning rate). Thus, we can use (2) to obtain a recursion on the new example sequence $\{\boldsymbol{\varphi}(n), d(n)\}$

$$\Omega(n+1) = \Omega(n) + \mu e^2(n)sign\{e(n)\}\varphi(n). \tag{3}$$

Repeating the application of (3), we obtain

$$\Omega(n+1) = \Omega(n-1) + \mu e^2(n-1)sign\{e(n-1)\}\varphi(n-1) \\ + \mu e^2(n)sign\{e(n)\}\varphi(n) \tag{4}$$

Rearranging (4), we have

$$\Omega(n+1) = \mu \sum_{j=1}^{n} \left[ e^2(j)sign\{e(j)\} \right] \varphi(j). \tag{5}$$

Here, $\varphi(n)$ is only implicitly known and its dimensionality is infinite for Gaussian kernel. For this reason, the derivation method of KLMS algorithm [7] is adopted for KLMAT algorithm. That is, compute the output of the filter $y(n+1)$ directly rather than expressing the weight vector. By using the *Mercer kernel*, the output of the filter can be calculated through kernel evaluations

$$y(n+1) = \Omega^T(n)\varphi(n+1) \\ = \mu \sum_{j=1}^{n} \left[ e^2(j)sign\{e(j)\} \right] \kappa(j, n+1) \tag{6}$$

where $\kappa(\mathbf{u}, \mathbf{u}') = \exp(-h \parallel \mathbf{u} - \mathbf{u}' \parallel^2)$ stands for *Gaussian kernel*, $h$ denotes the kernel size. The KLMAT algorithm adds a new space $e(n)sign\{e(n)\}$ for $\mathbf{u}(n+1)$ at each iteration, slightly increases the computational complexity as compared with the KLMS algorithm. We define $f_n$ as a nonlinear mapping at the $n$th iteration, the learning process of the KLMAT algorithm can be summarized as follows:

$$\begin{aligned} f_n &= \mu \sum_{j=1}^{n} \left[ e^2(j)sign\{e(j)\} \right] \kappa(\mathbf{u}(j), \cdot), \\ f_n(\mathbf{u}(n)) &= \mu \sum_{j=1}^{n} \left[ e^2(j)sign\{e(j)\} \right] \kappa(\mathbf{u}(j), \mathbf{u}(n+1)), \\ e(n+1) &= d(n+1) - f_n(\mathbf{u}(n+1)), \\ f_{n+1} &= f_n + \mu e^2(n+1)sign\{e(n+1)\}\kappa(\mathbf{u}(n+1), \cdot). \end{aligned} \tag{7}$$

For simplicity, the coefficient $a_j(n+1)$ is defined as:

$$a_j(n+1) = \mu \left[ e^2(j)sign\{e(j)\} \right], \quad j = 1, ..., n+1 \tag{8}$$

and $C(n)$ is referred as a center set at time $n$

$$\mathbf{C}(n+1) = [\mathbf{C}(n), \mathbf{u}(n+1)]. \tag{9}$$

The notation $C(n)$ is a codebook of the filter which stores the new center at each iteration.

### 2.2 VSS-KLMAT algorithm

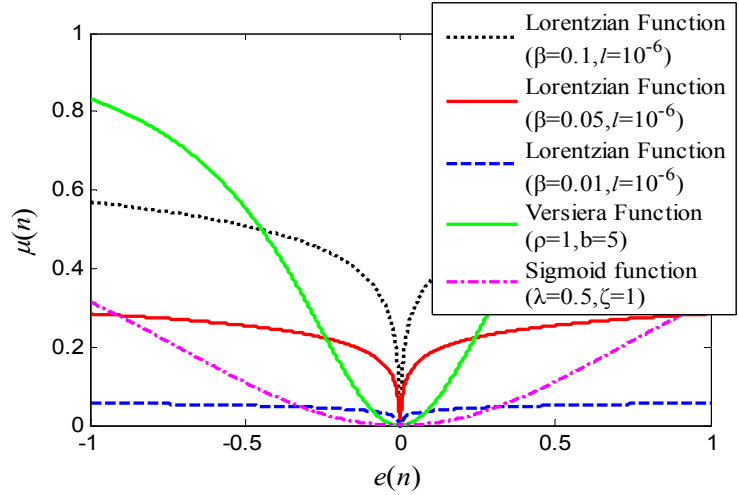

**Fig. 1.** Comparison of Lorentzian function, Versiera function and Sigmoid function.

An important limitation of the KLMAT algorithm is the convergence rate vs misadjustment trade-off imposed by the selection of a certain value for the step size. Motivated by the VSS scheme and *Lorentzian function* in [10], in this section, we proposed a novel VSS scheme for the KLMAT algorithm. Replacing $\mu$ with $\mu(n)$ for each iteration, $\mu(n)$ is adapted by using following expression

$$\mu(n) = \beta \log\left(1 + \frac{1}{2}\frac{e^2(n)}{l^2}\right) \qquad (10)$$

where $\beta$ is a scalar factor of the function which controls the value range of the function, and $l$ is the positive parameter. By using some nonlinear function, the VSS scheme has been developed in several previous studies, including *Sigmoid function* [11] and *Versiera function* [12]. It can be observed from Fig. 1 that the Lorentzian function is much steeper than the other functions for the same small error signal. Therefore, such function may achieve fast convergence rate and improved tracking capability when the unknown system changes abruptly in initial convergence stage.

To further improve the performance of the VSS scheme, an estimation of $e^2(n)$ is introduced to (10)

$$\delta_e(n+1) = \theta \delta_e(n) + (1-\theta)e^2(n) \qquad (11)$$

where $\theta$ is the forgetting factor that governs the averaging time constant, and $\delta_e(n+1)$ is a low-pass filtered estimation of $e^2(n)$. In stationary environments, the previous samples include information that is relevant to determining a measure of update, i.e., the proximity of the adaptive filter coefficients to the optimal ones. Hence, $\theta$ should be close to 1.

We set to $\theta=0.9$ for the VSS-KLMAT algorithm. Moreover, for the stability of the VSS strategy, $\mu(n)$ is further limited by

$$\mu(n+1) = \begin{cases} \mu_{max}, & \mu(n) > \mu_{max} \\ \mu_{min}, & \mu(n) < \mu_{min} \\ \mu(n), & otherwise \end{cases} \tag{12}$$

where $\mu_{max} = 2$ is the upper bound of the learning rate and $\mu_{min} = 0.01$ is the lower bound of the learning rate ( $0 < \mu_{min} < \mu_{max}$ ).

## 3 Analysis of the proposed algorithms

### 3.1 Convergence analysis of the proposed algorithm

To verify the stability of the proposed algorithms, the analysis of the convergence property is performed. The following assumptions are made to facilitate analysis:

1) $\boldsymbol{\varphi}(n)$, $d(n)$ and $\boldsymbol{\Omega}(n)$ are statistically stationary, and $\boldsymbol{\varphi}(n)$ and $d(n)$ are joint independent;

2) $\boldsymbol{\Omega}(n)$ is a zero mean with symmetrical distribution, $\sigma_v^2 \triangleq E\{\boldsymbol{\Omega}^2(n)\}$, $\varepsilon_v^4 \triangleq E\{\boldsymbol{\Omega}^4(n)\}$ and $\xi_v^6 \triangleq E\{\boldsymbol{\Omega}^6(n)\}$ exist.

3) $\boldsymbol{\varphi}(n)$ and $\boldsymbol{\Omega}(n)$ uniformly bounded, that is: $u \triangleq \sup_n \|\boldsymbol{\varphi}(n)\| < \infty$ and $\bar{\boldsymbol{\Omega}} = \sup_n |\boldsymbol{\Omega}(n)| < \infty$. For translation invariant kernel, the result value is a constant $\|\boldsymbol{\varphi}(n)\| = \sqrt{g_0}$, where $g_0$ is a constant.

4) $\boldsymbol{\varphi}(n)$ and $\boldsymbol{\Omega}(n)$ are statistically independent.

The weight deviation vector in feature space $\mathbf{V}(n)$ is defined as

$$\mathbf{V}(n) = \boldsymbol{\Omega}(n) - \boldsymbol{\Omega}_o . \tag{13}$$

where $\boldsymbol{\Omega}_o$ is the optimal weight vector. The adaptation of the KLMAT algorithm falls into the kind of recursive updates of the form

$$\mathbf{V}(n+1) = \mathbf{V}(n) + \mu F(\boldsymbol{\varphi}(n)) g(e(n)) \tag{14}$$

where $F(\boldsymbol{\varphi}(n)) = \boldsymbol{\varphi}(n)$ and

$$g(e(n)) = e(n)^2 sign(e(n)) . \tag{15}$$

Assuming the linearised system has no noise interference ($v(n)=0$), we define $\tilde{g}(n) = g(e(n))|_{v(n)=0}$, giving[1] [13]

$$\tilde{g}(n) = -(\boldsymbol{\varphi}^T(n)\mathbf{V}(n))^2 sign(\boldsymbol{\varphi}^T(n)\mathbf{V}(n)). \tag{16}$$

Let $\mathbf{V}_*$ be a stable stationary point on the error surface, the Jacobian of the $F(\boldsymbol{\varphi}(n))g(e(n))$ is obtained by using $sign(x) = x/|x|$

$$\begin{aligned}\mathbf{B}_{\mathbf{V}_*}(n) &= F(\boldsymbol{\varphi}(n))\frac{\partial \tilde{g}^T(n)}{\partial \mathbf{V}(n)}|_{\mathbf{V}(n)=\mathbf{V}_*} \\ &= -\left[\frac{3(\boldsymbol{\varphi}^T(n)\mathbf{V}_*)^2|\boldsymbol{\varphi}^T(n)\mathbf{V}_*|}{|\boldsymbol{\varphi}^T(n)\mathbf{V}_*|^2} - \frac{|\boldsymbol{\varphi}^T(n)|(\boldsymbol{\varphi}^T(n)\mathbf{V}_*)^2\mathbf{V}_*}{|\boldsymbol{\varphi}^T(n)\mathbf{V}_*|^2}\right]\boldsymbol{\varphi}(n)\boldsymbol{\varphi}^T(n).\end{aligned} \tag{17}$$

Note that

$$c(n,\boldsymbol{\varphi},\mathbf{V}_*) = \frac{3(\boldsymbol{\varphi}^T(n)\mathbf{V}_*)^2|\boldsymbol{\varphi}^T(n)\mathbf{V}_*| - |\boldsymbol{\varphi}^T(n)|(\boldsymbol{\varphi}^T(n)\mathbf{V}_*)^2\mathbf{V}_*}{|\boldsymbol{\varphi}^T(n)\mathbf{V}_*|^2}. \tag{18}$$

Thus, the linearised system in $\mathbf{V}_*$ is

$$\begin{aligned}\mathbf{V}(n+1) &= \left[\mathbf{I} - \mu\mathbf{B}_{\mathbf{V}_*}(n)\right]\mathbf{V}(n) \\ &= \left[\mathbf{I} - \mu c(n,\boldsymbol{\varphi},\mathbf{V}_*)\boldsymbol{\varphi}^T(n)\boldsymbol{\varphi}(n)\right]\mathbf{V}(n)\end{aligned} \tag{19}$$

For $\mathbf{V}_* = 0$, $c(n,\boldsymbol{\varphi},\mathbf{V}_*) = 0$. By using the total stability theorem (TST), the assumption $v(n)=0$ is removed [14]. The TST can prove the exponential stability of the adaptive system by using a linearised system. The ordinary difference equation is expressed as [14]:

$$\mathbf{V}(n+1) = \mathbf{A}(n)\mathbf{V}(n) + \mathbf{H}(n,\mathbf{V}(n)) + \mathbf{L}(n,\mathbf{V}(n)) \tag{20}$$

where $\mathbf{A}(n)$ is the transition matrix of the linearised system, $\mathbf{A}(n)\mathbf{V}(n)$ is a linearised system, $\mathbf{H}(n,\mathbf{V}(n))$ and $\mathbf{L}(n,\mathbf{V}(n))$ are the difference between (14) and (19). Introducing (12), (13) and (17) to (18), we arrive

$$\begin{aligned}&\mathbf{H}(n,\mathbf{V}(n)) + \mathbf{L}(n,\mathbf{V}(n)) \\ &= \mu\left[g(e(n))\boldsymbol{\varphi}(n) - c(n,\boldsymbol{\varphi},\mathbf{0})\boldsymbol{\varphi}(n)\boldsymbol{\varphi}^T(n)\mathbf{V}(n)\right]\end{aligned} \tag{21}$$

In KLMAT algorithm, $\mathbf{H}(n,\mathbf{V}(n))$ and $\mathbf{L}(n,\mathbf{V}(n))$ can be identified as follows:

$$\mathbf{H}(n,\mathbf{V}(n)) = \mu\left[e^2(n)sign(e(n))\boldsymbol{\varphi}(n)\right] \tag{22}$$

$$\mathbf{L}(n,\mathbf{V}(n)) = \mathbf{0} \tag{23}$$

The four *regularity conditions* of TST [14] can be expressed by

$$\begin{cases} I : \mathbf{H}(n,\mathbf{0}) = \mathbf{0} \\ II : \| \mathbf{H}(n,\mathbf{V}_1) - \mathbf{H}(n,\mathbf{V}_2) \| \leq \partial_1 \| \mathbf{V}_1 - \mathbf{V}_2 \| \\ III : \| \mathbf{L}(n,\mathbf{V}) \| \leq \partial_2 \hbar \\ IV : \| \mathbf{L}(n,\mathbf{V}_1) - \mathbf{L}(n,\mathbf{V}_2) \| \leq \partial_2 \| \mathbf{V}_1 - \mathbf{V}_2 \| \end{cases} \quad (24)$$

where $\partial_1$ and $\partial_2$ are the positive constants, the weight error vector is range from the region $\Lambda$, $\Lambda = \{\mathbf{V} : \| \mathbf{V} \| < \hbar, \hbar > 0\}$ [15]. If $n \geq 0$, $(\mathbf{V}, \mathbf{V}_1, \mathbf{V}_2) \in \Lambda$.

---

[1] The linearised system has assume to have no noise inference $v(n)=0$, which is based on the Linearisation theory [14]. This assumption does not agree with the practical adaptive systems, but it laid the groundwork for total stability theorem (TST).

Clearly, the above-mentioned regularity conditions are completely satisfied via (21) and (22). Therefore, it can be concluded that the KLMAT algorithm is an exponential stability under the systematic noise.

### 3.2 Convergence analysis of the mean weight vector

To guarantee the stability of the KLMAT algorithm, the choice of the step-size $\mu$ would be discussed.

Substituting (13) into (5) and taking the mathematical expectation of both sides, we have

$$E(\mathbf{V}(n+1)) = E(\mathbf{V}(n)) - \mu E\left(e^2(n) sign(e(n))\boldsymbol{\varphi}(n)\right). \quad (25)$$

By using the Price's theorem [17], the second term of the right hand side in (25) is simplified as:

$$E\left(e^2(n) sign(e(n))\boldsymbol{\varphi}(n)\right) \approx \frac{2\sqrt{2}\sigma_e(n)\mathbf{R}_{\varphi\varphi}}{\pi} E(\mathbf{V}(n)) \quad (26)$$

where $\mathbf{R}_{\varphi\varphi} = E(\boldsymbol{\varphi}^T(n)\boldsymbol{\varphi}(n))$ is the auto-correlation matrix of input vector in feature space, $\sigma_e(n)$ is the root mean-squared estimation error at time $n$. Combining (25) and (26), we get

$$E(\mathbf{V}(n+1)) = \left(1 - \mu \frac{2\sqrt{2}\sigma_e(n)\mathbf{R}_{\varphi\varphi}}{\pi}\right) E(\mathbf{V}(n)). \quad (27)$$

The mean weight vector is converged if and only if

$$\prod_{n=0}^{L} \left(1 - \mu \frac{2\sqrt{2}\sigma_e(n)\mathbf{R}_{\varphi\varphi}}{\pi}\right) \to 0, \text{ as } L \to \infty \quad (28)$$

Thus, a sufficient condition to ensure the convergence of the mean weight vector is obtained:

$$0 < \mu < \sqrt{\frac{\pi}{2}} \frac{1}{\sigma_e(n)\lambda_{\max}} \quad (29)$$

where $\lambda_{\max}$ is the maximum eigenvalue of the matrix $\mathbf{R}_{\varphi\varphi}$.

### 3.3 Analysis of the parameter *l*

The choice of parameter *l* affects the performance of the VSS-KLMAT algorithm. By employing the analyzed result of mean weight vector, the range of parameter *l* can be analyzed. Combining (10) and (29), we obtain

$$\log\left(1+\frac{1}{2}\frac{e^2(n)}{l}\right) < \sqrt{\frac{\pi}{2}}\frac{1}{\sigma_e(n)\lambda_{\max}\beta}. \tag{30}$$

Using the Taylor series, the left term of the (30) can be approximately given as

$$\log\left(1+\frac{1}{2}\frac{e^2(n)}{l}\right) = \frac{\ln\left(1+\frac{1}{2}\frac{e^2(n)}{l}\right)}{\ln 10} \approx \frac{e^2(n)}{2\ln 10\, l}. \tag{31}$$

Then the range of *l* can be obtained by

$$l > \frac{e^2(n)\sigma_e(n)\lambda_{\max}\beta}{\sqrt{2\pi}\ln 10}. \tag{32}$$

Obviously, *l* should be greater than zero for stability.

### 3.4 Computational complexity

Table 1 Computational complexity.

| Algorithms | Computation (training) | Memory (training) | Computation (test) | Memory (test) | Computation Time(seconds) |
|---|---|---|---|---|---|
| LMAT | $O(N)$ | $O(M)$ | $O(M)$ | $O(M)$ | 0.060548 |
| KLMS | $O(N^2)$ | $O(N)$ | $O(N)$ | $O(N)$ | 15.554524 |
| **KLMAT** | $O(N^2)$ | $O(N)$ | **$O(N)$** | **$O(N)$** | **20.511481** |
| **VSS-KLMAT** | $O(N^2)$ | $O(N)$ | **$O(N)$** | **$O(N)$** | **21.441392** |

Table 1 summarizes the computational complexity of the algorithms, where *N* is the training times and *M* is the length of the kernel filter. As can be seen, the kernel adaptive algorithms require more operations as compared to the linear algorithms, the increase of the kernel adaptive algorithms in computational complexity can be compromised by their excellent nonlinear signal processing capability. With additional 1 multiplication for computing coefficient $a_j(n+1)$, the KLMAT algorithm behaves much better than the KLMS algorithm under the different noise environments. The VSS-KLMAT algorithm is developed based on VSS scheme, which requires additional 7 production for computing coefficient $a_j(n+1)$. Thus, its computation complexity slightly increases.

To quantify the computational burden, we measured the average run execution time of the algorithm on a 2.1-GHz AMD processor with 2GB of RAM, running Matlab R2013a on Windows 7 environment. As one see from Table 1, the LMAT algorithm is the fastest method owing to utilizing gradient descent. The VSS-KLMAT algorithm increases the computation time, but the faster initial convergence rate is achieved as compared to other algorithms.

## 4  Simulation results

We conduct a series of simulations to evaluate the performance of the proposed algorithms, including simulations on a Mackey–Glass (MG) chaotic time series prediction and simulations on a sunspot number time series analysis. We compare the estimation results of the proposed algorithms with those of the LMAT algorithm and the KLMS algorithm. In the simulation study, the effectiveness is assessed in terms of mean square error (MSE) in testing stage, which is defined as $\mathrm{MSE} = 10\log_{10}\{e^2(n)\}$. The parameters in algorithms (learning rate, kernel size, etc.) are selected to guarantee the fast and stable convergence of the algorithms. All the simulation results below are averaged over 100 independent Monte Carlo runs.

### 4.1  Example 1

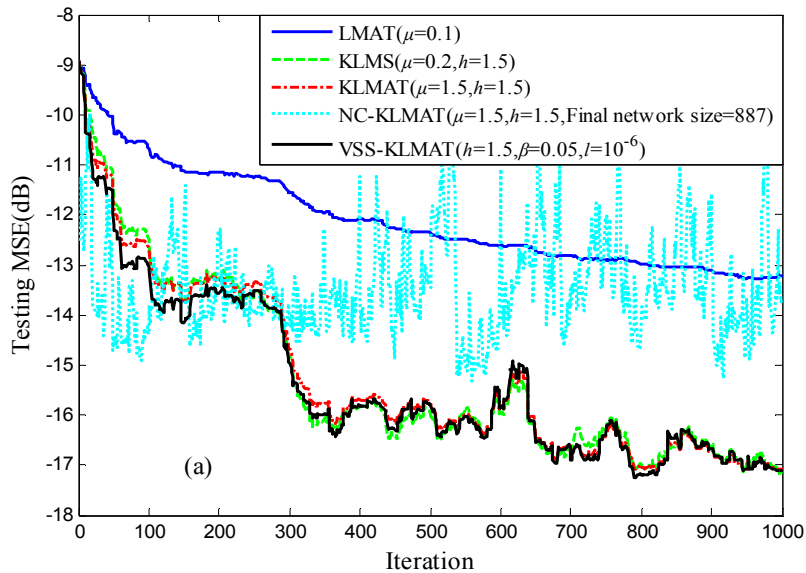

In the first example, simulation studies are carried out for the MG chaotic time series prediction. The MG series is generated by a delay ordinary differential equation [7]:

$$\frac{dx(t)}{dt} = -qx(t) + \frac{mx(t-\tau)}{1+x(t-\tau)^{10}} \tag{33}$$

where $q$=0.1, $m$=0.2, and $\tau$=30. The sampling period is 6 seconds and time embedding (filter order) is 10. The White Gaussian noise (WGN) and impulsive noise $v(n)$ are used in this simulation. The WGN with zero mean and standard deviation $\sigma_G$ is added to the chaotic time series prediction. The impulsive noise is modelled by Bernoulli-Gaussian (BG) distribution with probability function $P_c$ and the deviation $\sigma_I^2$ [1,18]. A segment of 1000 samples is used as the training data and another 1000 as the test data.

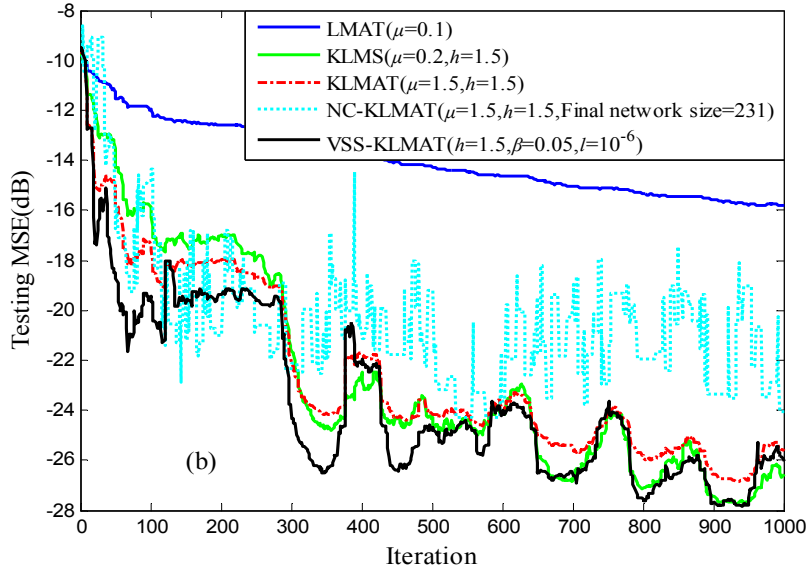

**Fig. 2.** Testing MSE curves of algorithms (a) WGN with $\sigma_G = 0.1$; (b) WGN and impulsive noise with $P_c$=0.3 and $\sigma_I = \sigma_G = 0.02$.

The size of the kernel network increases by one kernel evaluation with every new training sample. This may prohibit the practical implementation of these algorithms for a large training set. Therefore, the novelty criterion (NC) strategy is considered to be a possible solution to the limitation of the KLMAT algorithm. The NC-KLMAT algorithm can be easily derived by introducing NC [4] in the KLMAT algorithm. Fig. 2 illustrates the proposed algorithms in time series prediction with WGN and impulsive noise environments. In the simulation, we compare the performance of five algorithms: LMAT, KLMS, NC-KLMAT, KLMAT and VSS-KLMAT. As can be seen, the KLMAT algorithm slightly increases the convergence rate as compared to KLMS algorithm in both cases. Besides, owing to utilizing the Lorentzian function, the proposed VSS-KLMAT algorithm achieves faster convergence rate than the KLMAT algorithm in both environments. The NC method provides overwhelmingly smaller network size as comparison to other algorithms with sacrifice of the steady-state performance.

### 4.2 Example 2

In the second example, we focus on the performance of the proposed algorithms in MG time series prediction for various probability densities. The experimental conditions are considered to be same as in the example 1, but the measurement noise is considered as non-Gaussian noise. Since in many physical environments, the noise is characterized by non-Gaussian distribution. Fig. 3 (a)-(d) illustrates a comparison of the LMAT, KLMS, KLMAT and VSS-KLMAT algorithms. We can clearly see that the KLMAT and VSS-KLMAT algorithms outperform the KLMS and LMAT algorithms for all of the probability densities with the same steady-state error. Specifically, the performance of the proposed algorithms is much better than that of the KLMS algorithm with Rayleigh noise and exponential noise.

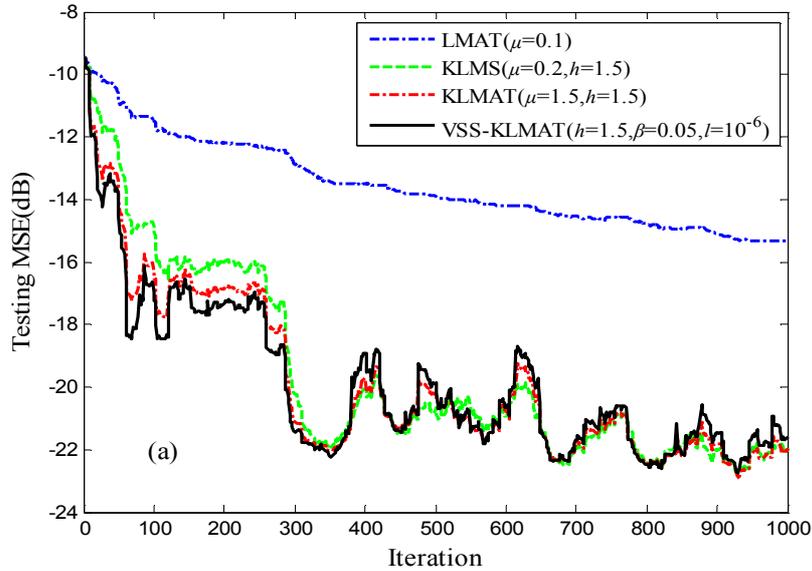

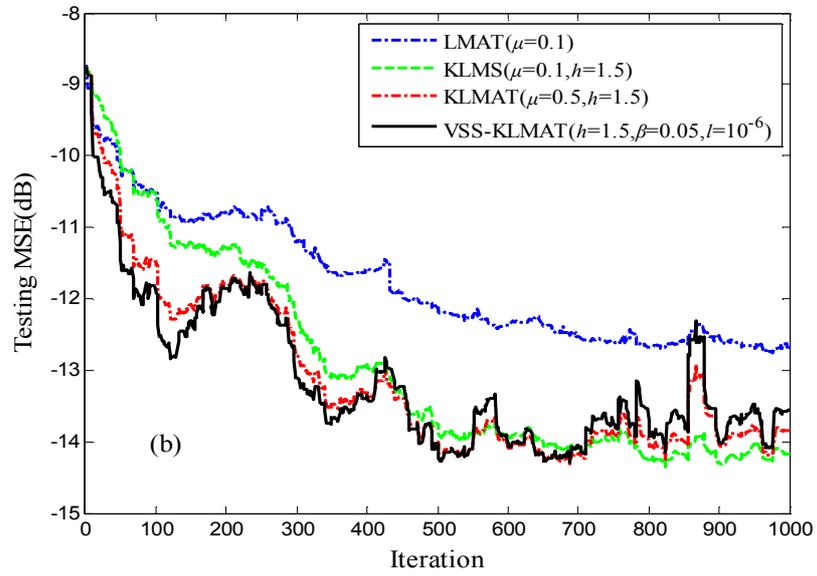

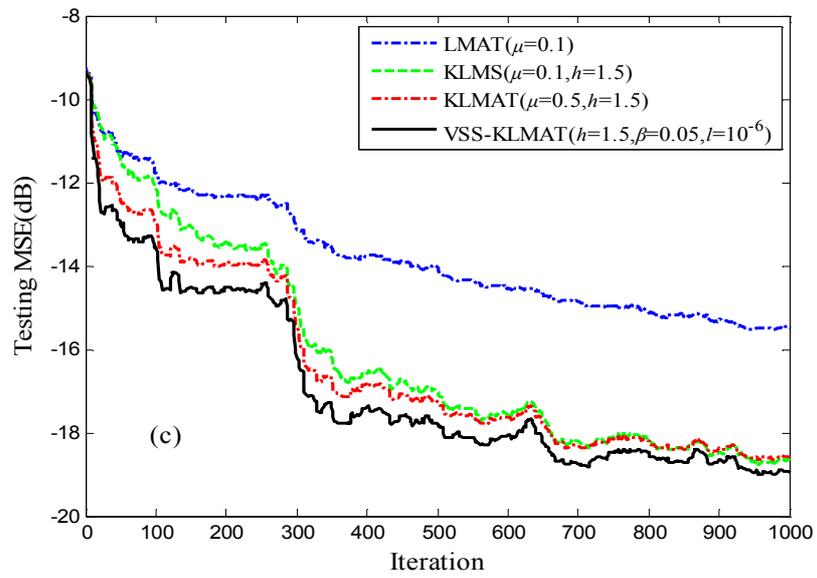

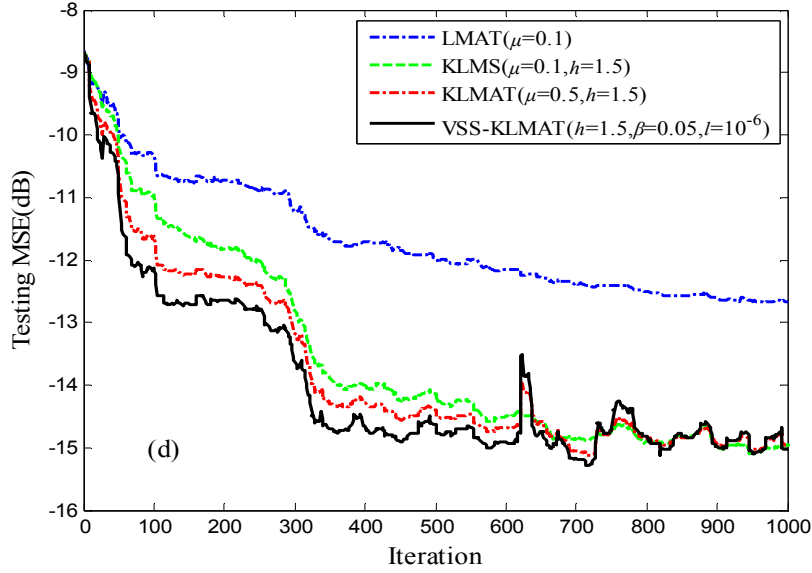

**Fig. 3.** MSE curves of algorithms in Mackey–Glass time series prediction for various probability densities. (a) Uniform noise with the variance $\sigma^2=5$; (b) Rayleigh noise $\sigma^2=0.05$; (c) Rectangular noise with amplitude $A=0.1$, frequency $f=1000$Hz; (d) Exponential noise with characteristic parameter 0.1.

### 4.3 Application of the Sunspot Number Time Series Analysis

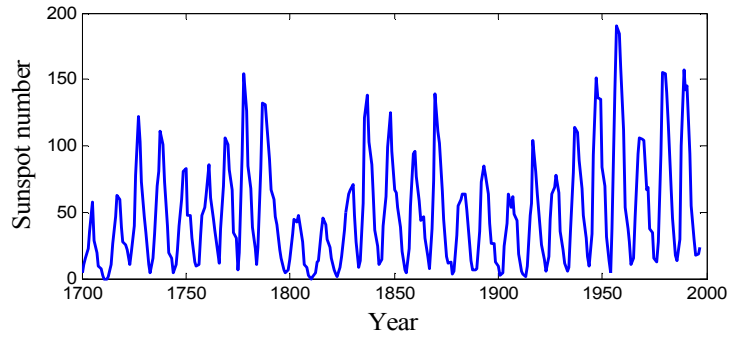

**Fig. 4.** Sunspot number time series.

To test the performance in realistic applications, the proposed algorithms are applied to adaptive prediction of the annually recorded sunspot time series for the years 1700–1997 [16]. In Fig. 4, the actual output of the sunspot number series is depicted. The time embedding is 2. A Gaussian kernel with kernel width 1.5 is chosen for all kernel adaptive filters. In Fig. 5(a), the prediction results for WGN with $\sigma_G = 0.1$ are illustrated. It is concluded that the proposed algo-rithms are superior to the LMAT and KLMS algorithms in terms of the convergence speed. Additionally, the prediction results with uniform noise are shown in Fig. 5(b). Again, faster convergence speed can be obtained by the proposed algorithms.

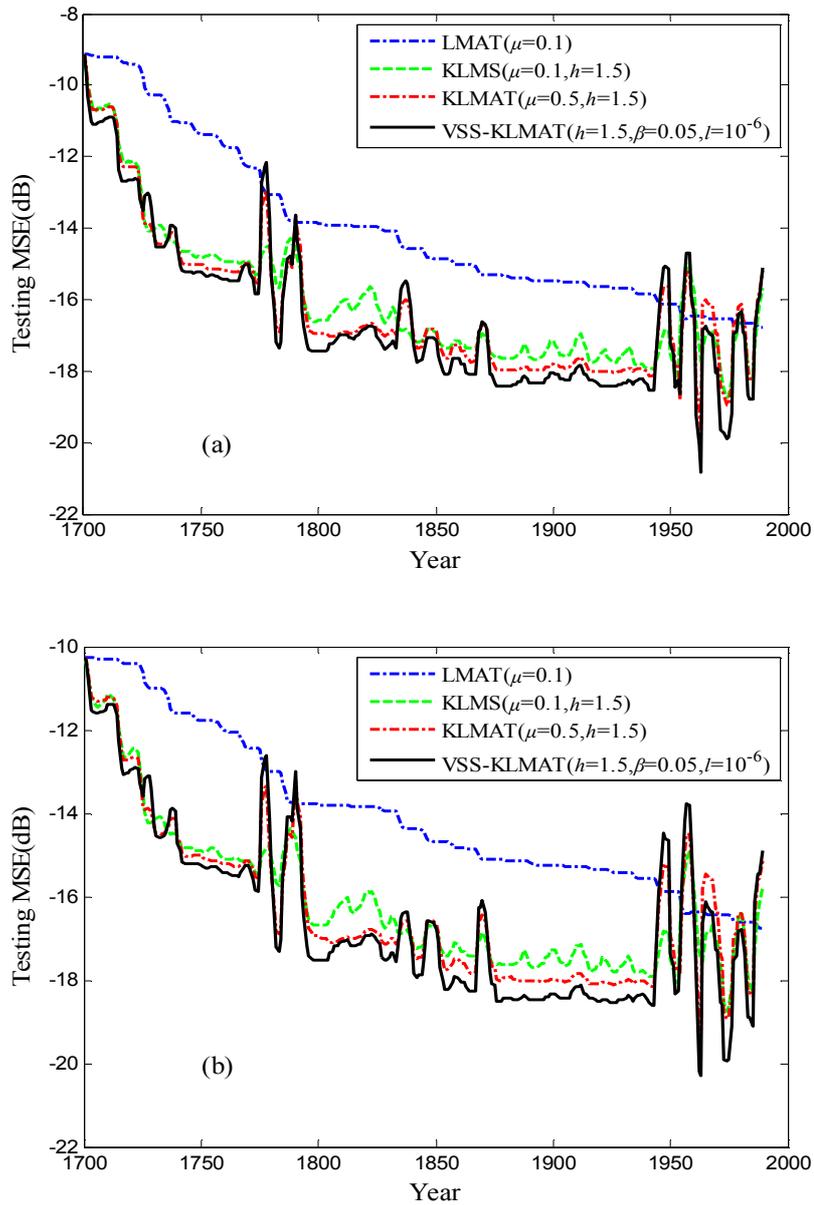

**Fig. 5.** Predicting results for sunspot time series. (a) WGN with $\sigma_G = 0.1$. (b) Uniform noise with $\sigma_G = 10$.

# 5 Conclusion

A KLMAT algorithm and its VSS version have been proposed by combining the LMAT algorithm with the kernel methods. The proposed algorithms have been shown to exhibit higher convergence rate than the KLMS algorithm for the same steady-state error. The exponential stability of the proposed algorithms is shown by application of the deterministic averaging analysis and the total stability theorem. We further provide simulation results that illustrate the efficiency of the proposed algorithms for time series prediction in different scenarios.

## Acknowledgments

This work was partially supported by National Science Foundation of P.R. China (Grant: 61571374, 61271340, 61433011). The first author would like to acknowledge the China Scholarship Council (CSC) for providing him with financial support to study abroad (No. 201607000050).